Postmortem analysis and possible rebirth of LK-99

Itai Panas

Department of Chemistry and Chemical Engineering

Chalmers University of Technology

S-412 96 Gothenburg, Sweden

Abstract

The LK-99 hype came and went but the great potential of the apatite class of materials as platform for flat bands' research must not be swept away in the process. A heuristic reinterpretation of the atomic structure of LK-99 is offered, including electronic signatures from the fully oxidized to the fully reduced. Here, copper is proposed to reside in the apatite channel rather than doping the Lead sublattice. Contact is made with the experimental x-ray powder diffractogram. The electronic signatures are found to reflect those of local [O-Cu-O] and [O-Cu-$V_O$**] molecular ions, where $V_O$** is a vacant oxygen site in the apatite channel. The local nature warrants flat bands. Charge carrier concentration is controlled by the oxygen content. Fully reduced LK-99 is a wide band gap insulator with a well-developed intermediate band. Fully oxidized LK-99 is a magnetic insulator, while the partially reduced LK-99 allows for hopping of holes between π* orbitals of non-magnetic and magnetic [O-Cu-O] moieties. The findings are extended to include the case where half of the Cu atoms are replaced by Ni and half by Zn. For [O-Ni-O] +[O-Zn-$V_O$**] inter-site accidental near degeneracy among states at $E_F$ is reported, while for [O-Zn-O] + [O-Ni-$V_O$**] the near-degeneracy at EF on hole doping is among intra-site valence states exclusively on [O-Ni-$V_O$**]. The apatite platform is ideal for the study of flat bands associated phenomena. The reinterpreted fully reduced LK-99 system invites intrinsic intermediate band solar cells applications. Sites with variable oxygen occupation allow for holes doping, the consequences of which include strong correlations between electrons and lattice that suggest the emergence of (rebuilt) phonons mediated intra- as well as inter- bands (virtual) electron transfer-based phenomena, feeding the quest for ambient conditions superconductivity.

Recent reports of ambient conditions superconductivity (ACS) sparked hectic activity in the scientific community during a couple of months before it was put out. The candidate was a Lead Copper Oxy-apatite compound $Pb_{10-x}Cu_x(PO_4)_6O$; $0.9<x<1.0$ [1,2], coined LK-99. Attempts to reproduce the ACS where either negative or inconclusive until the hypothesis was put forward that the effect was an artefact possibly due to a $Cu_2S$ impurity. The ACS in LK-99 was subsequently refuted by means of experiment [3,4].

While order was restored in the experimental community, the theoretical electronic structure echelon suffered. Confidence in the ability of electronic structure calculations to differentiate between a wide band gap insulator and a potential superconductor needs rebuilding also by showing how they may in fact be connected.

In as much as understanding of materials properties starts with viable atomic structures, acquaintance must be made with the system at hand regarding viable modifications. For some of us, the LK-99 hype served as an eye opener regarding the variety and versatility the apatite class of materials [5-9], and crucially, to materials that can readily be synthesized and modified. Consider the biological bone-forming mineral $Ca_5(PO_4)_3OH$ as starting point for reinterpreting LK-99. Replacement of $H^+$ by $M^+$ is suggested to lead to $Ca_5(PO_4)_3OH_{1-x}M_x$. Indeed, extensive studies of $A_5(PO_4)_3OH_{1-x}M_x$, A=Ca, Sr, Ba, tell of M residing in the apatite channels and forming, e.g. $[O-M-O]^{2-}$, $[M-O]^-$, $[O-M-O]^{3-}$ molecular ions for M=Fe, Co, Ni, Cu, Zn. It is gratifying to note that such molecular entities have been reported in the literature [10-12]. It is also noted that if M(II) then we obtain $Ca_5(PO_4)_3OH_{1-x}M_{x/2}$. By taking A=Pb, M=Cu(II) and x=1, the composition $Pb_5(PO_4)_3OCu_{0.5}$ is arrived at, that may be extended to allow for variable oxygen content, i.e. $Pb_{10}(PO_4)_6[O-Cu(II)-O]$ (fully oxidized), $Pb_{20}(PO_4)_{12}[O-Cu(II)-V_O^{**}][O-Cu(II)-O]$ (fully reduced), and $Pb_{30}(PO_4)_{18}[O-Cu(II)-V_O^{**}][O-Cu(II)-O]_2$ (partially reduced), where $V_O^{**}$ is an +2 charged oxygen vacancy situated in the center of the $Pb_3$ sites that form the apatite channel on stacking, see Figure 1. Given that the corresponding stoichiometries are $Pb_{10}P_6O_{25}Cu$, $Pb_{10}P_6O_{25.5}Cu$, and $Pb_{10}P_6O_{26}Cu$, let the sum of Cu and Pb atoms to be 10, i.e., 10(Pb:10/11 + Cu:1/11), then we obtain $Pb_{10-x}Cu_xP_6O_{25+y}$, y=0, 0.5 or 1 and, x=0.9. In fact, in what follows, we will assume that these compounds jointly comprise the corresponding LK-99 compounds.

**Figure 1.** Lead copper oxy-apatite where we have Cu to reside in the apatite channel rather than doping the Pb sublattice as assumed in most studies of LK-99. (**A**) Horizontal view. (**B**) Top view. (**C**) Orbital characteristics emphasizing the "lone-pair" on Pb(II) and the triangular $Pb_3$ $V_O^{**}$ site.

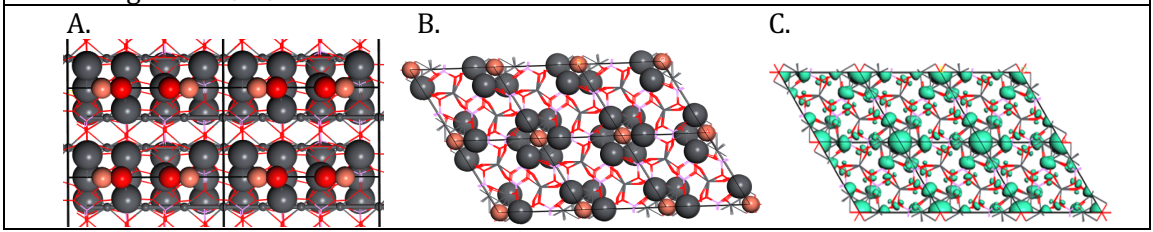

Various superstructures of each of fully oxidized, fully reduced, and partially reduced LK-99 were considered. Starting out with the nominal $Pb_5(PO_4)_3Cu_{0.5}O$ $P6_3/m$ space group of apatite, they were subjected to full geometry optimization employing the PBESOL GGA functional [13] as implemented in CASTEP [14]. Ultrasoft pseudopotentials [15] with a 571 eV cut-off energy were used in conjunction with Monkhorst Pack k-points grids with grid parameters less than 0.045 Å$^{-1}$. PBE0 hybrid functional [16] calculations were resorted to, to quantify the band gaps in the fully reduced LK-99. Powder X-ray diffractograms (PXRDs) were simulated using the REFLEX software [17] and employed to compare with experiment [18]. In what follows, selected generic features are presented for each of the three LK-99 systems. Moreover, replacement of Cu by Ni/Zn is undertaken to demonstrate the claimed versatility.

The atomic structure of $Pb_{10}(PO_4)_6[O-Cu(II)-O]$ that obeys the P-3 space group is considered here as a representative of fully oxidized LK-99. Calculations tell of a molecular $[OCu(II)O]^{2-}$ anion, where in particular Cu $3d_{xz}$ and $3d_{yz}$ mix with O $2p_x$ and $2p_y$ to form six 3-center molecular orbitals of π symmetry to accommodate eleven electrons: four electrons enter the bonding π orbital, four electrons the non-bonding π orbital, and three into the anti-bonding π orbital, i.e., $\pi^{*3}$, compare Figures 2 and 3.

| **Figure 2.** Molecular orbitals characteristics. ||
|---|---|
| **A**. Selected orbitals of $[O-Cu(II)-O]^{x-}$ anions. | **B**. Selected orbitals of $[Cu(II)O]^-$ anions. |
| Non-bonding δ–orbitals | Non-bonding δ–orbitals |
| 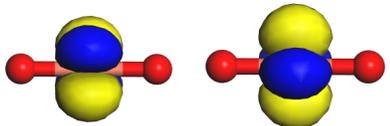 | 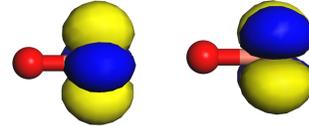 |
| Doubly degenerate bonding π–orbital | Doubly degenerate anti-bonding π–orbital |
| 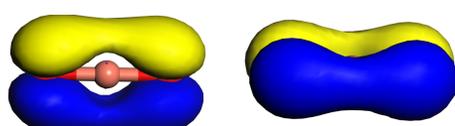 | 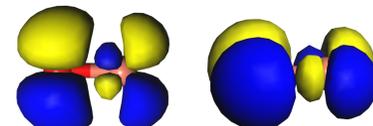 |
| Bonding and non-bonding σ–orbital | Bonding σ–orbital |
| 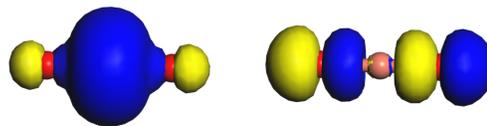 | 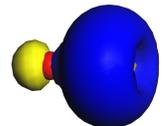 |
| Doubly degenerate anti-bonding π–orbital | Anti-bonding σ–orbital |
| 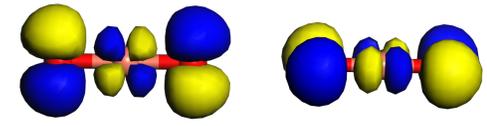 | 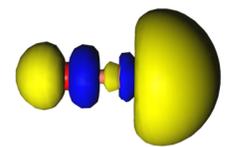 |
| • Acceptor in $\pi^{*3}$ $[OCu(II)O]^{2-}$<br>• Donor in filled $\pi^{*4}$ $[OCu(II)O]^{3-}$ | Unoccupied in $[O-Cu(II)-V_O^{**}]^-$ |

**Figure 3.** Characteristics of fully oxidized LK-99, that is $Pb_{10}(PO_4)_6[O-Cu(II)-O]$. **(A)** Comparison of experimental and computed powder x-ray diffractograms a=9.906 Å, c=7.446Å. Spin density maps: top- **(B)** and side- **(C)** views on the apatite channel when occupied by $[O-Cu(II)-O]^{2-}$ molecular anion emphasizing the magnetic $\pi^{*3}$ electronic structure. **(D)** α(blue) and β(red) spin density of states.

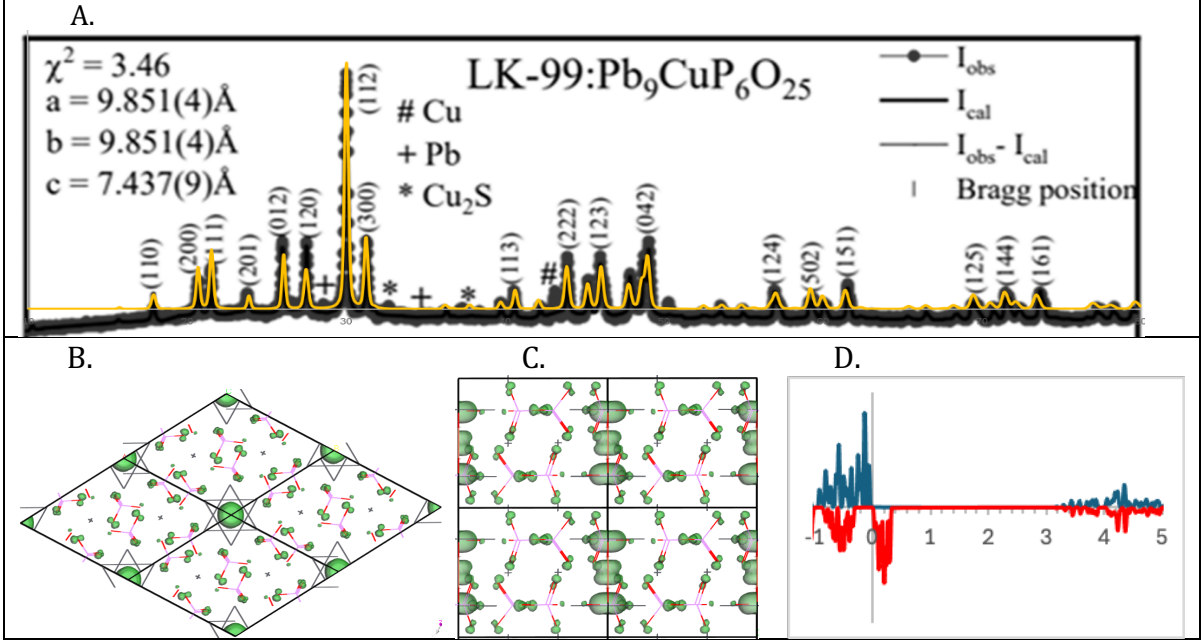

In as much as $[O-Cu(II)-O]^{2-}$ is magnetic, so is also the fully oxidized LK-99. Throughout, we make connection to the PXRD pattern of LK-99 in the literature to support the assumed structural relevance, see Figure 3 again.

To produce the fully reduced moiety we first double the unit cell to obtain $Pb_{20}(PO_4)_{12}[O-Cu(II)-O][O-Cu(II)-O]$ and then remove an oxygen atom to arrive at $Pb_{20}(PO_4)_{12}[O-Cu(II)-V_O^{**}][O-Cu(II)-O]$. One of the two electrons thus released render the $\pi^*$ orbital in the resulting $[O-Cu(II)-O]^{3-}$ filled, i.e., $\pi^{*4}$, while the second electron renders the $\pi^*$ orbital associated with $[O-Cu(II)-V_O^{**}]^-$ molecular anion filled, leaving empty the $\sigma^*$ hybrid orbital that mixes 4s, 4p, $3d_{3z^2-r^2}$ on Cu, see Figure 2 again. Satisfactory agreement with the experimental PXRD was obtained for the DFT optimized atomic structure, see Figure 4. The basic electronic structure signature of this fully reduced LK-99 candidate is flat bands in line with the ionic nature of the material. The top of the valence band (TVB) is dominated by the filled [O-Cu-O] $\pi^{*4}$ orbital, while PBESOL GGA has the position of an intermediate band (IB) associated with the $\sigma^{*0}$ $[O-Cu-V_O^{**}]$ at +1.3 eV. Correspondingly, the bottom of the conduction band (BCB) is at +2.8 eV. These values for the PBE0 hybrid functional come out at +3.0 eV and +4.8 eV for the IB and BCB, respectively. Neither the GGA nor the hybrid functional predict anything but insulator behavior. However, when inspecting the fully reduced LK-99 we note the distinct (photo-) excitation that is required to occupy the IB, i.e., from $\pi^{*4} [O-Cu(II)-O]^{3-} + \sigma^{*0} [O-Cu(II)-V_O^{**}]^- => \pi^{*3} [O-Cu(II)-O]^{2-} + \sigma^{*1} [O-Cu(II)-V_O^{*}]^{2-}$. We note further the high DOS beyond the BCB that suggests utilizing this property in intrinsic intermediate band solar cell [19] applications, see Figure 4 again.

**Figure 4.** Characteristics of fully reduced LK-99 $Pb_{20}(PO_4)_{12}[O-Cu(II)-V_O^{**}][O-Cu(II)-O]$. (**A**) Comparison of experimental (black) [8] and our theoretical (orange) PXR diffractograms from DFT optomized strcuture Projected cell parameters: a=9.87; b=9.85 Å; c=14.81 Å (7.405 Å); (**B**) Atomic structure of $Pb_{20}(PO_4)_{12}[O-Cu(II)-O][Cu(II)-O]$ straight/bent from DFT. Left: Phosphates made transparent for clarity. Right: Crystal orbitals: Blue is TVB; Green is IB. (**C**) Left: PBESOL GGA DOS; Center: PBESOL GGA Band structure, TVB (blue), IB (green). Right: PBE0 DOS(eV). Total DOS(eV) (dotted), PDOS(eV) $[OCuV_o^{**}]$ (red)), PDOS(eV) [OCuO] (green), -PDOS(eV)$[V_o^{**}=Pb_3]$(black). (**D**) Ferromagnetic excited state Left: spin density. Right: spin DOS(eV). α(blue) and β(red).

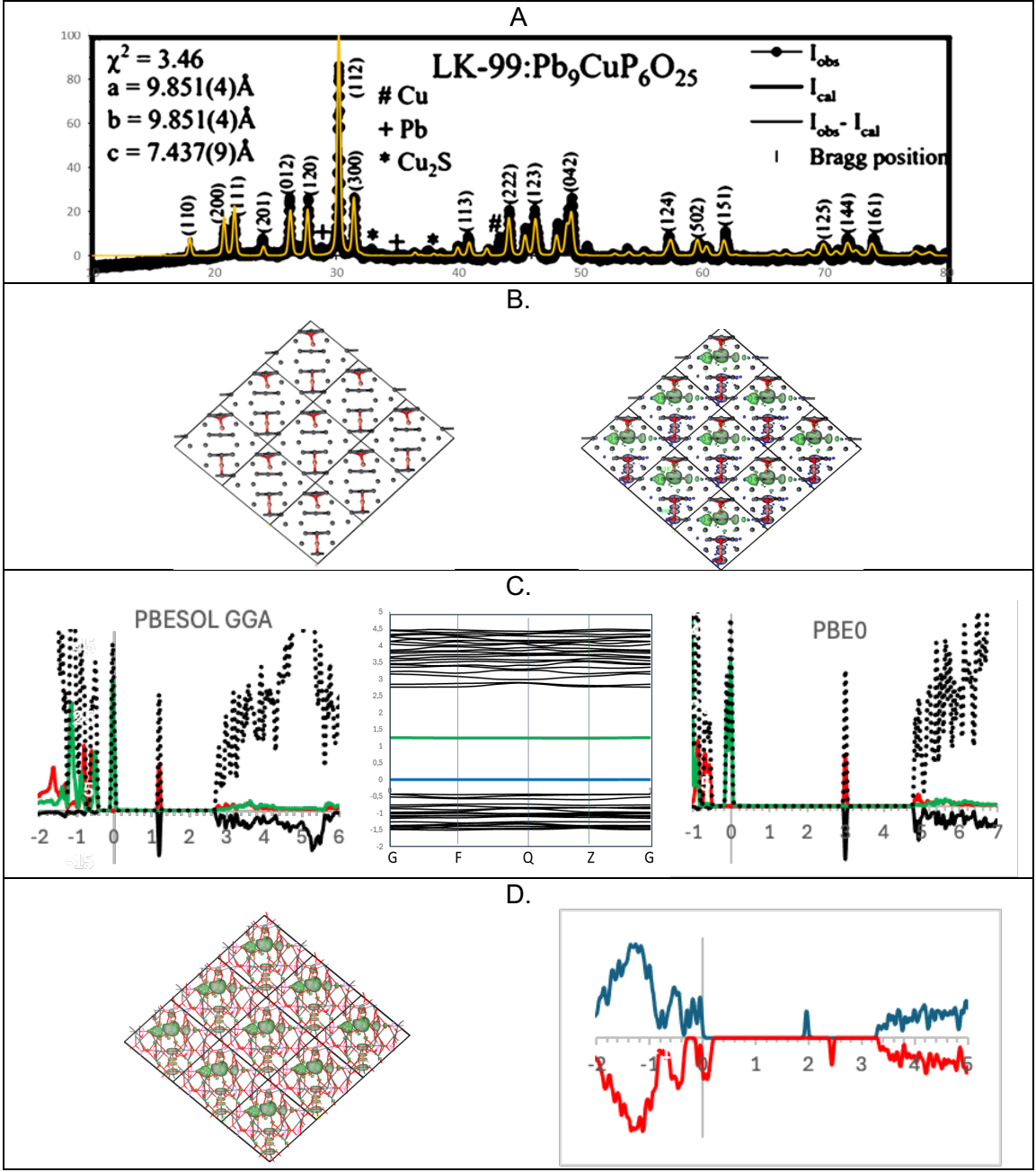

It is not inconceivable, however, that the LK-99 is only partially reduced. To capture this aspect here, we consider $Pb_{30}(PO_4)_{18}[O-Cu(II)-V_O^{**}][O-Cu(II)-O]_2$ for which one candidate structure in the P3 space group is provided in Figure 5. The $\sigma^{*0}$ [O-Cu-$V_O^{**}$] IB that appears because of the reduction is pushed away from the TVB if compared to the fully reduced LK-99, cf. Figures 4 and 5. This, owing to a hole doping of the fully reduced LK-99. Note that formally [O-Cu(II)-$V_O^{**}$]$^-$ is accompanied by $\pi^{*3}$ [O-Cu(II)-O]$^{2-}$, and $\pi^{*4}$ [O-Cu(II)-O]$^{3-}$. The self-interaction error (SIE) is understood to disrupt the strong correlation between electrons and the lattice, to produce two equivalent 3-center linear [O-Cu(II)-O]$^{2.5-}$ molecular ions, each with approximately 3.5 electrons in their $\pi^*$ orbitals. In reality, it is not inconceivable that the lack of overlap between [O-Cu(II)-O] moieties, combined with inherent degeneracy, leads to spontaneous symmetry breaking into [O-Cu(II)-O]$^{2-}$ and [O-Cu(II)-O]$^{3-}$, and possible phonon assisted (virtual) shuffling of localized electrons and holes.

**Figure 5.** Characteristics of partially reduced LK-99, $Pb_{30}(PO_4)_{18}[O-Cu(II)-V_O^{**}][O-Cu(II)-O]_2$. (**A**) Comparison of experimental (black) [8] and our theoretical (orange) PXR diffractograms from DFT optomized strcuture Projected cell parameters: P3 a:9.912 Å, c=22.357 Å (7.452 Å). (**B**) Ferromagnetic state Left: Electronic spin density. Right: spin DOS(eV). α(blue) and β(red).

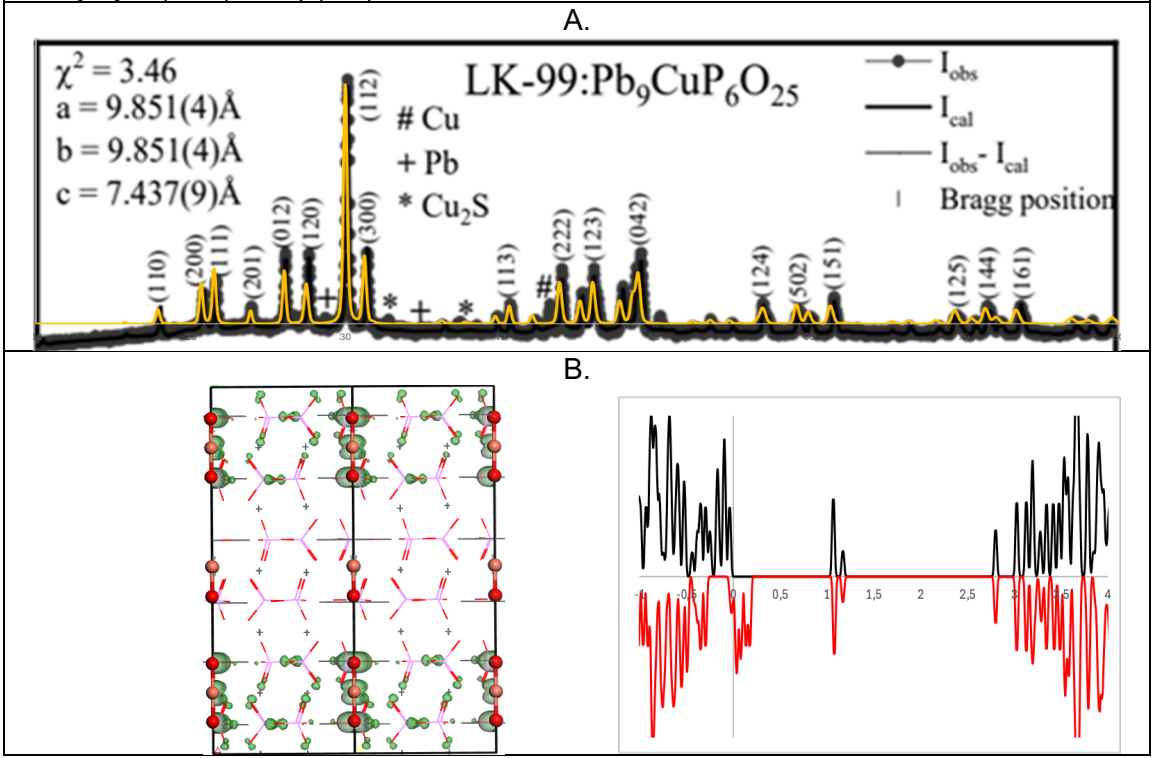

Besides possible photo-assisted electron transfer from TVB to IB in the fully reduced LK-99, it becomes interesting to explore chemical means to close the TVB-IB gap. Noting that 2Cu is iso-electronic with Zn + Ni, consider first the corresponding fully oxidized $Pb_{20}(PO_4)_{12}$[O-Zn(II)-O][O-Ni(II)-O] system that has $\pi^{*4}$ [O-Zn(II)-O]$^{2-}$ and $\pi^{*2}$ [O-Ni(II)-O]$^{2-}$, the latter being Hund's rule stabilized. We arrive at a fully developed IB either by removing an oxygen from O-Zn(II)-O or from O-Ni(II)-O, see Figure 6. We achieve accidental near-degeneracy of the incommensurate IB and TVB co-existing at the Fermi energy when we have TVB $\pi^{*3}$ [O-Ni(II)-O]$^{3-}$ + IB $\sigma^{*1}$ [O-Zn(II)-$V_o^*$]$^-$. It should be noted that the TVB-IB gap opens up if we instead have $Pb_{20}(PO_4)_{12}$[O-Ni(II)-$V_o^{**}$][O-Zn(II)-O] corresponding to IB $\sigma^{*0}$ [O-Ni(II)-$V_o^{**}$]$^0$ + TVB $\pi^{*4}$ [O-Zn(II)-O]$^{2-}$. The latter restores the generic features of the original fully reduced LK-99. However, it is noted that on hole doping the latter, by annihilating some of the oxygen vacancies, the holes hopping takes place among remaining $\pi^{*4}$ [O-Ni(II)-$V_o^{**}$]$^0$ and $\pi^{*3}$ [O-Ni(II)-$V_o^{**}$]$^+$ sites, while in all-Cu LK-99 the holes reside on the [O-Cu(II)-O] anions. Thus, the presumed (virtual) holes hopping would be mediated by different phonons in the two systems.

**Figure 6.** Characteristics of $Pb_{20}(PO_4)_{12}$[O-Ni(II)-O][O-Zn(II)-$V_o^*$] (Left column) and $Pb_{20}(PO_4)_{12}$[O-Zn(II)-O][O-Ni(II)-$V_o^{**}$] (right column) (**A**) Crystal orbitals at $E_F$. (**B**) Total DOS(eV) (dotted), PDOS(eV)[$OM_1V_o^{**}$] (red)), PDOS(eV)[$OM_2O$] (green), -PDOS(eV)[$V_o^{**}$<=>$Pb_3$](black). Left: $M_1$=Zn; $M_2$=Ni. Right: $M_1$=Ni; $M_2$=Zn. (**C**) B enlarged.

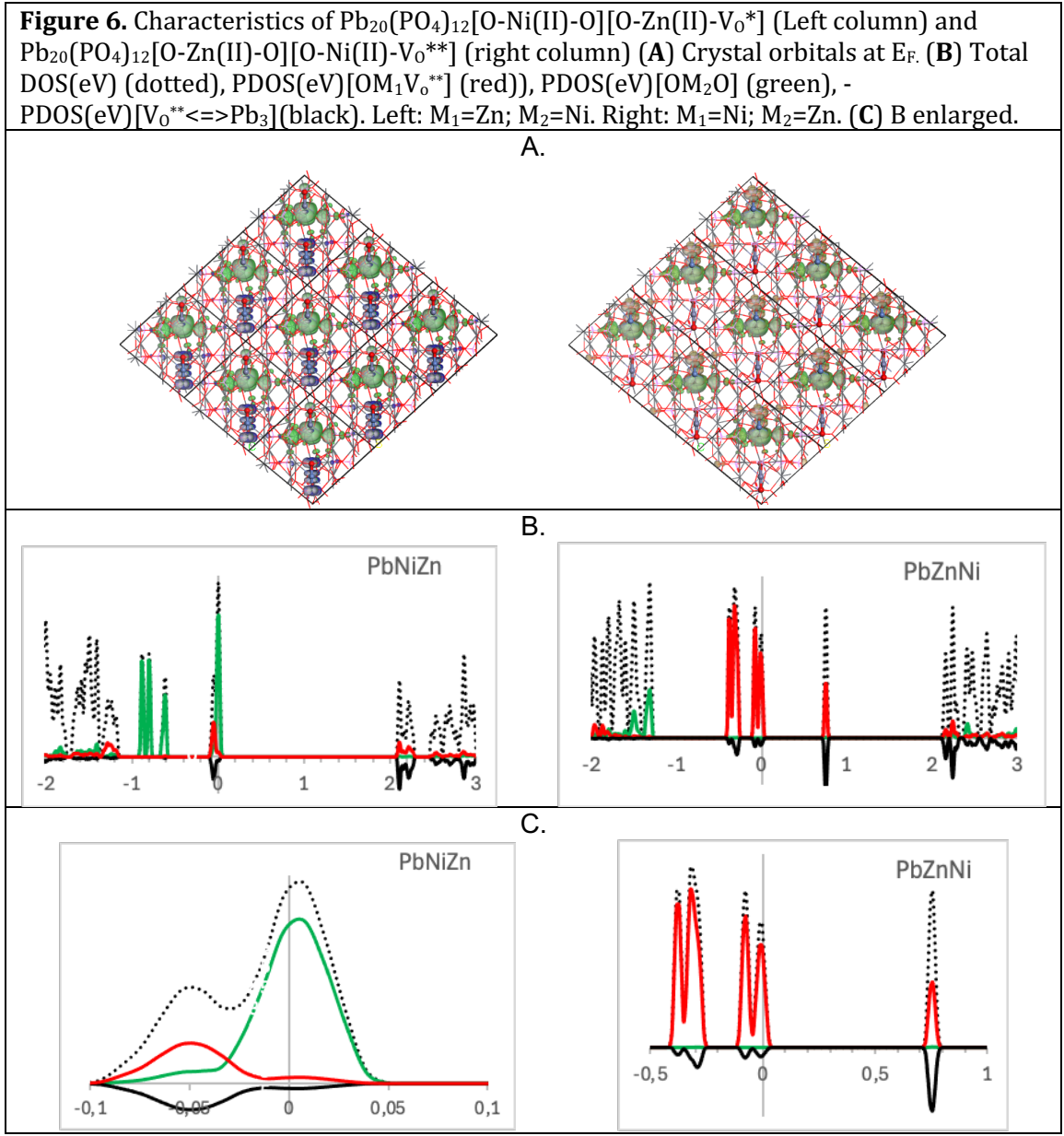

To conclude, we have reinterpreted the crystal structure of LK-99 and clarified basic electronic features of the material to result from Cu situated in the apatite channel rather than doping in the Pb sublattice. Structural characteristics as captured by experimental PXRD compare reasonably well to those obtained here by DFT for the fully oxidized, fully reduced and partially reduced atomic structure models of LK-99.

- For the fully oxidized LK-99, the electronic properties are decided by the magnetic $\pi^{*3}$ [O-Cu(II)-O]$^{2-}$ molecular anions that reside in the apatite channel.
- For the fully reduced LK-99, the states associated with the band gap belong to $\pi^{*4}$ [O-Cu(II)-O]$^{3-}$ (TVB), and $\sigma^{*0}$ [O-Cu(II)-V$_O$**]$^{-}$ (IB) molecular anions.
- Partially reduced LK-99 supports the notion of holes hopping between $\pi^{*3}$ [O-Cu(II)-O]$^{2-}$ and $\pi^{*4}$ [O-Cu(II)-O]$^{3-}$.
- On substituting Cu for 50%Ni and 50%Zn in fully reduced LK-99 accidental degeneracy of incommensurate bands at EF is observed, originating from $\pi^{*3}$ [O-Ni(II)-O]$^{3-}$ + $\sigma^{*1}$ [O-Zn(II)-V$_O$*]$^{-}$ associated states.
- Characteristics of all-Cu fully reduced LK-99 are somewhat recovered for the switched $\pi^{*4}$ [O-Zn(II)-O]$^{4-}$ + $\sigma^{*0}$ [O-Ni(II)-V$_O$**]$^{0}$ arrangement. The IB is dominated by the said $\sigma^{*0}$ orbital on [O-Ni(II)-V$_O$**]$^{0}$, while the TVB is composed of the underlying $\pi^{*}$ orbitals on the same moiety.
- Hole hopping in partially oxidized LK-99(Zn,Ni) takes place among the residual p*4 [O-Ni(II)-VO**]0 and p*3 [O-Ni(II)-VO**]+ sites that remain unoccupied by oxygen.

The versatile apatite platform [20] is ideal for the study of flat bands associated phenomena. The reinterpreted fully reduced LK-99 system invites intrinsic intermediate band solar cells applications. Sites with variable oxygen occupation allow for holes doping, the consequences of which include strong correlations between electrons and lattice that suggest the emergence of (rebuilt) phonons mediated intra- as well as inter- bands (virtual) electron transfer-based phenomena, all feeding the quest for ambient conditions superconductivity hitherto not achieved [21].